\documentclass[letterpaper]{article} % DO NOT CHANGE THIS
\usepackage[]{aaai25}  % DO NOT CHANGE THIS
\usepackage{times}  % DO NOT CHANGE THIS
\usepackage{helvet}  % DO NOT CHANGE THIS
\usepackage{courier}  % DO NOT CHANGE THIS
\usepackage[hyphens]{url}  % DO NOT CHANGE THIS
\usepackage{graphicx} % DO NOT CHANGE THIS
\usepackage{natbib}  % DO NOT CHANGE THIS AND DO NOT ADD ANY OPTIONS TO IT
\usepackage{caption} % DO NOT CHANGE THIS AND DO NOT ADD ANY OPTIONS TO IT
\usepackage{colortbl}
\usepackage{multirow}
\usepackage{multicol}
\usepackage{tabularx}
\usepackage{xcolor}
\usepackage{afterpage} % Crucial for clean layout transitions
\usepackage{lipsum}
\usepackage{booktabs}
\usepackage{makecell}
\usepackage{array}
\usepackage{longtable}
\usepackage[export]{adjustbox}

\usepackage{algorithm}
\usepackage{algorithmic}

\usepackage{newfloat}
\usepackage{listings}
\DeclareCaptionStyle{ruled}{labelfont=normalfont,labelsep=colon,strut=off} % DO NOT CHANGE THIS
\floatstyle{ruled}
\newfloat{listing}{tb}{lst}{}
\floatname{listing}{Listing}
\title{Shaping the Future of Generative AI for Black Communities: A Frame Analysis of Public Discourse and Empirical Scholarly Research}
\author{
    Angela D. R. Smith\textsuperscript{\rm 1},
    Gabriella Thompson\textsuperscript{\rm 2}, Christopher L. Dancy\textsuperscript{\rm 3}\\ Mark Diaz \textsuperscript{\rm 4}, Seyi Olojo\textsuperscript{\rm 5}, Christina N. Harrington \textsuperscript{\rm 4}
}

\affiliations{
    \textsuperscript{\rm 1}University of Rochester\\
    \textsuperscript{\rm 2}University of Texas at Austin\\
    \textsuperscript{\rm 3}The Pennsylvania State University\\
    \textsuperscript{\rm 4}Google Research\\
    \textsuperscript{\rm 5}University of California\\ Berkeley\\
    a.smith@rochester.edu
}

\usepackage{bibentry}
\begin{document}

\maketitle

\begin{abstract}
As generative AI (genAI) systems become embedded in education, employment, healthcare, and creative industries, the impact and engagement among marginalized groups have become both a widespread discourse and a focus in scholarly research. As a starting point, we examine public discourse and empirical research to explore the impact of genAI systems on Black communities. We conducted a systematic literature review (SLR) of 91 empirical papers alongside a media discourse frame analysis of 28 public resources, applying Entman's framing theory to map how each corpus defines problems, attributes causes, and proposes treatments. Our SLR reveals that scholarly research concentrates heavily on technical bias detection, reducing Blackness to measurable variables rather than engaging with cultural practices, structural conditions, or Black knowledge systems. Our frame analysis reveals that public discourse attributes genAI-related harm to historical and systemic forces, while scholarly research stops its causal accounts at the dataset and its treatment recommendations at technical reform. We demonstrate that this misalignment is structurally produced: anti-Blackness operates simultaneously across both registers, generating a shared evacuation of Black epistemic agency. We argue for frame analysis as an AI ethics methodology capable of surfacing what technical evaluation forecloses.
\end{abstract}

\section{Introduction}

Public discourse shapes how societies understand, adopt, and regulate emerging technologies. As generative AI (genAI) systems become increasingly embedded in education, employment, healthcare, and creative industries, narratives circulating in media outlets, policy briefs, and opinion pieces play a crucial role in signaling anticipated harms, assigning responsibility, and proposing policy and technical interventions. For historically marginalized communities, particularly Black Americans who have borne disproportionate harms from algorithmic systems ranging from biased facial recognition to discriminatory hiring tools~\citep{noble2018algorithms, benjamin2019race}, understanding how public discourse frames genAI's potential impacts is essential for informing research agendas, policy priorities, and community organizing efforts. Similarly, empirical research and the translational impact observed in research papers have begun to directly influence the policy, design, development, and use of genAI tools~\cite{septiandri2024impact}. Such scientific explorations are especially pivotal to addressing existing and anticipated harms, underscoring the value of the knowledge cultivated by scholarly research. 

This surfaces a critical question that remains underexplored: How do the concerns and framings present in public discourse about genAI and Black communities align with—or diverge from—the priorities of research that seeks to directly study and engage with these communities? As this work evolves, it is also essential to interrogate conceptions of Blackness and \textit{how} Black communities are engaged and the ways Black culture and identity are portrayed to inform AI development. Considering such interactions requires us to interrogate the extractive nature of AI research and development \cite{network2025technoskepticism}, and move beyond the oppressive nature of ``inherently biased algorithms''. If public discourse emphasizes the importance of community-driven approaches to identifying and mitigating structural harms such as cultural erasure (i.e., when AI corrects for bias by censoring parts of Black history \cite{Small_2023}), scholarly research may miss the mark of potential impact by focusing predominantly on automating the evaluation of technical metrics such as bias detection. This misalignment may indicate that research is failing to address the issues that matter most to affected communities and the broader public, while also flattening Blackness to merely be categorical dimensions that typically do not capture the full human condition of this community. 

This paper investigates this potential gap through a frame analysis of public discourse concerning genAI and Black communities. Drawing on Entman's~\citeyearpar{entman1993framing} widely-used framing theory, we systematically analyze media resources—including news articles, opinion pieces, and policy briefs—to identify how public narratives define problems, attribute causes, render moral judgments, and propose treatments related to genAI's impact on Black Americans. We complement this analysis with a systematic literature review (SLR) of empirical literature from research institutions published at computing venues to characterize how research has approached genAI and Black communities, providing context for assessing alignment between public discourse and scholarly priorities.

Our analysis reveals significant divergences between public discourse and empirical scholarly research. Using framing theory signals and emphasizes several dimensions of harm that can be attributed to systemic bias and marginalization, while the SLR suggests a heavy research focus on technical bias detection. We offer three contributions to the AIES community. First, we identify and characterize the dominant frames shaping public narratives about genAI's anticipated impacts on Black communities. Second, we demonstrate the misalignment between public discourse and empirical research, identifying it as not incidental but structurally produced. Third, we argue for frame analysis as an AI ethics methodology — one that captures the structural and anticipatory dimensions of harm that individual-focused technical evaluation forecloses, and that makes visible what research epistemology is structured to exclude. In doing so, we respond directly to AIES's remit of interrogating the societal dimensions of AI: frame analysis extends the community's methodological repertoire beyond system-level evaluation toward the discursive and structural conditions under which AI harms are defined, contested, and governed.

We argue that attending to public discourse offers valuable insights for AI ethics scholarship. Public frames reveal what harms are anticipated before they fully materialize, what causal explanations resonate with broader audiences, and what interventions are seen as viable—information that can inform both research agendas and policy development. By mapping the terrain of public concern and research focus, this paper contributes to ongoing efforts to ensure that AI research addresses the issues that matter to the communities most likely to be affected by these powerful technologies.

\section{Blackness, Racialization \& Relation to GenAI}
Understanding the relationship between genAI and Black communities requires situating contemporary bias within the historical processes of racialization that shape how AI systems are developed, deployed, and experienced. While Black people, communities, and practices trace origins well beyond recent forms of racialization, current racial structures significantly influence technology at micro (individual identities and interpersonal interactions), meso (organizational and institutional practices), and macro (societal structures, law, and policy) levels~\cite{ray19}. The logics underlying race and Blackness from Western, white perspectives, what Wynter and McKittrick~\cite{wynterm15} describe as the ``standard genre of the human'', are embedded in the engineering lifecycle of AI systems~\cite{workman26inlets}, producing the specific manifestations of bias documented in empirical research.

These racialized logics have evolved from theological to biological to cultural rationales, but consistently position Blackness as inferior or deviant~\cite{robinson20, kendi17, wynter03}, operating through what Workman and Dancy~\cite{workman26inlets} identify as ``inlets'' where anti-Blackness enters AI development at every level. Within Human-Computer Interaction (HCI), this manifests through ``solutionist'' approaches that treat Blackness as a problem to be corrected rather than recognizing the ontological complexity of Black experience and knowledge~\cite{cunningham23}. Colorblind perspectives~\cite{bonillasilva15} and the fluidity of racial categories further obscure the persistence of these logics, making structural intervention difficult even as Black communities continue to imagine with and beyond racialization~\cite{brayh21}.

\section{Related Work}
\subsection{Field Goals and Motivation of GenAI}
The values structuring AI development shape how systems are built and assessed, with consequences for minoritized communities. \citet{birhane2022values} identify that AI scholarship prioritizes quantitative empiricism, efficiency, and performance over societal need and negative impacts — a pattern reflected in the field's reliance on benchmark datasets, which have been shown to value efficiency and universality over care and contextuality~\cite{scheuerman2021datasets, qadri2025case}. These field-level values extend to fairness paradigms: ~\citet{hanna2020towards} argue that dominant approaches treat oppressed identities as interchangeable, lacking race-consciousness, while Blodgett et al.~\cite{blodgett2020language} call for researchers to account for power relations between themselves and the communities they study.

\subsection{Biases and Harms in GenAI}
Scholarly research has documented a range of social biases in genAI systems, from racial bias in language models~\cite{sap2019risk} to gender bias in image generation~\cite{wan2024survey}, generating frameworks for identifying and mitigating model bias~\cite{diaz2024sound}. Shelby et al.~\cite{shelby2023sociotechnical} offer the most comprehensive taxonomy of sociotechnical algorithmic harms — representational, allocative, quality of service, interpersonal, and social system — which we adopt in our analysis. Research specifically focused on Black communities has identified failures to understand African American communication~\cite{brewer2023envisioning}, inadequate support for Black cultural contexts~\cite{olojo2025lost}, and reproduction of cultural stereotypes in genAI outputs~\cite{hofmann2024ai}. Despite the value of this work, researchers have called for new evaluation methods commensurate with genAI's scale and scope, including methods rooted in HCI~\cite{weidinger2023sociotechnical}.

\subsection{Bridging Black Communities and AI Development}
Public perceptions of AI influence how systems are developed and deployed~\cite{lima2023blaming, liu2024understanding}, and these perceptions vary across race, class, and gender~\cite{kelley2021exciting}. Black communities hold complex, often critical attitudes toward AI — particularly in healthcare contexts~\cite{lee2021included, parker2026} — and report pessimism about algorithmic fairness and bias in online platforms~\cite{harris2023honestly}. These perceptions have motivated calls for responsible inclusion of Black communities in AI development, including feminist-reflexive data curation~\cite{leavy2021ethical}, critical race theory for AI fairness~\cite{hanna2020towards}, and CRT-grounded HCI~\cite{ogbonnaya2020critical, smith2020s}. Participatory approaches have demonstrated particular value: co-design with Black adults for mental health agents~\cite{oleary2025}, speculative design with Black youth for justice-oriented AI principles~\cite{tanksley2025ethics}, and community workshops for equitable speech technology~\cite{brewer2023envisioning}. Yet scholars note that participation alone is insufficient without intentional structural commitment~\cite{sloane2022participation, birhane2022power}. Our work contributes to this area by examining the parallel terrain of public discourse and empirical research to map where emphases, priorities, and concerns align and where they diverge.

\section{Methods}
To map this landscape, we employed a dual-methodological approach combining a systematic literature review (SLR) with a frame analysis of discourse found in public media and reports. Contrasting public concerns and priorities with focus areas dominating empirical research helps to reveal misalignments that may indicate where research is failing to address community-identified needs and anticipatory harms. We detail our process below (see also Figure \ref{fig:methods}).

\begin{figure*}
    %\centering
    %\hspace*{-.2cm}
    \includegraphics[width=2.1\columnwidth]{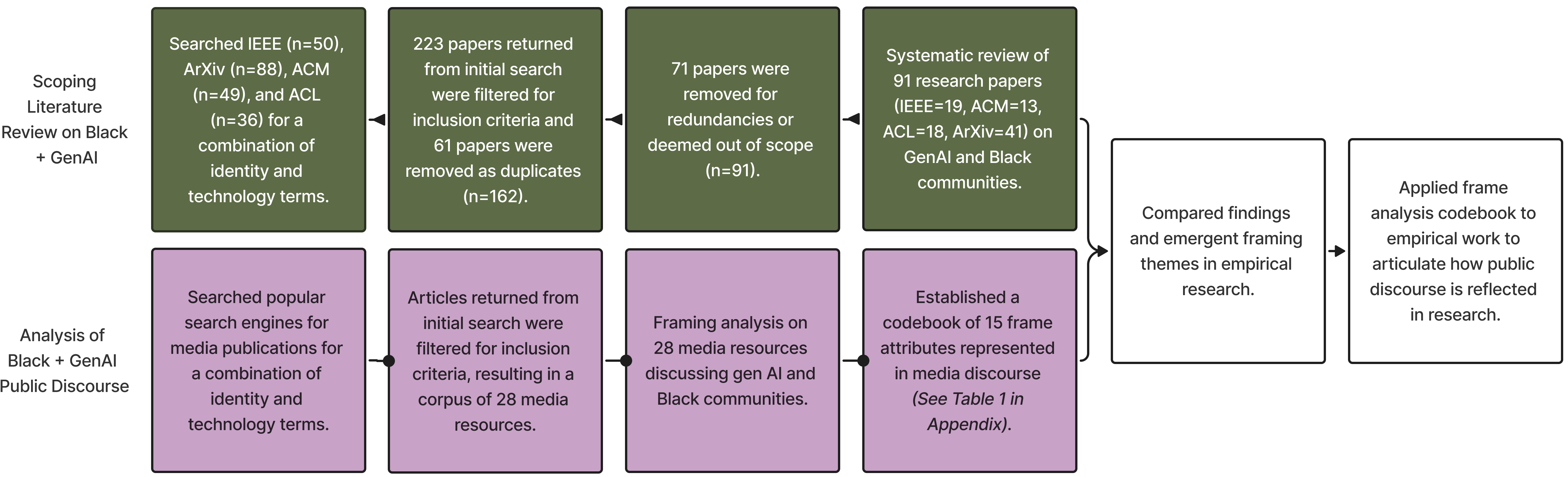}
    \caption{Systematic Review and Framing Analysis flow diagram.}
    \label{fig:methods}
\end{figure*}

\subsection{Systematic Literature Review}
\subsubsection{Data Collection and Corpus Creation}
To capture the landscape of research on genAI conducted with Black communities, we conducted a systematic literature review of full research papers published in major computing and AI databases between January 2010 and December 2025. We sought to assess the focus and priorities of existing research, to examine how Black communities were engaged in the ideation, evaluation, and testing of genAI tools, and to identify existing research on their perceptions, attitudes, and opinions of genAI~\cite{haimson2025}. We focused on publications from the ACM (encompassing HCI venues including CHI, CSCW, DIS, FAccT, and TOCHI) and IEEE digital libraries, and specialized AI venues, such as ACL, EAAMO, and NeurIPS.\footnote{We initially included AAAI in our database search; however, in trying to repeat some of the searches, we found the search system itself produced inconsistent results across repeated identical queries, undermining the reproducibility our protocol required. As such, we removed those papers from our corpus and note this as a coverage limitation, given AAAI's prominence as an AI/ML venue.}\textsuperscript{,}\footnote{Scopus was excluded due to its substantial overlaps with the ACM and IEEE digital libraries. Google Scholar was considered unsuitable as a principal search system for a systematic review due to its non-transparent ranking, personalized results, and unstable retrieval~\cite{gusenbauer2020academic}.} We also searched ArXiv, a venue where several prominent AI papers have been published to date. While other venues in the social sciences and humanities may have been relevant to the conversation of understanding the reception of genAI among various historically marginalized communities, we prioritized papers published in the computing field. 

Three members of the research team defined search terms within two categories: \textit{``solo/group racial identity terms''} and \textit{``generative AI terms''}. \textit{Solo racial identity terms} included `Black communities', `Black populations', or `Black Americans'. \textit{Group racial identity terms} included terms like `people of color', `marginalized/minoritized communities', or `vulnerable populations'. Our search also incorporated all possible generative AI terms including `large language models', `text-to-text', , `ChatGPT', `AI image generation', and `prompt engineering' and derivations of these terms. 

We then searched using a combination of terms from each category, for example: `Black people' + `generative AI', or `African American' + `text-to-image'. We filtered research articles with the following inclusion criteria: 1) Must focus on genAI (images, text, and/or videos); 2) Must be a full peer-reviewed paper published between 2010-2025; 3) Be published in English; 4) Must be empirical research where Black population is the focus of a qualitative study (interviewing, focus groups, participatory design), quantitative study (testing and evaluation of systems, surveys), or computational techniques (dataset and model evaluation frameworks). Works in progress, conference extended abstracts, and poster submissions were omitted. 

All authors iterated on a survey coding instrument and survey items to ensure mutual understanding during coding. We chose five papers to test the coding instrument for reliability and validity among researchers. Once the coding instrument was finalized, the remaining papers in the larger corpus were split and coded by one of four researchers.

Our search returned 223 papers across the five databases. Four members of the research team completed a title and abstract search to remove papers that did not fit the purpose of this review (n=71), and any duplicate papers (n=61) that may have returned over multiple databases. We then removed 18 papers that did not focus on Black communities but marginalized groups more broadly, leaving 91 papers for coding and analysis (19 IEEE, 41 ArXiv, 13 ACM, 18 ACL) (Figure \ref{fig:methods}).  

\subsubsection{Analysis}
Our coding instrument was used to analytically process dimensions and concepts and deductively extract data from 91 research articles. Papers were coded for year of publication, venue, search terms present in the paper, type of research study, type of genAI, intended contribution, domain or context of genAI application, whether the paper focused on an artifact or dataset, whether the artifact or dataset was evaluated, and forms of engagement. We also assessed whether the authors defined Blackness or other terms of identity. We used an aggregate analysis of closed-ended questions to describe trends in publication timelines and venues, the frequency of paper topics by type of genAI system, and research types. Open-ended questions such as ‘Contribution’ and ‘Purpose/Goal’ were analyzed thematically. After our initial coding of the paper corpus, we noticed that our papers largely centered harms and harm mitigation. As such, we recoded our corpus using the frame attributes, notably~\citet{shelby2023sociotechnical}'s sociotechnical harms, which helped us articulate the harms present in the papers.  

\subsection{Frame Analysis}
We use frame analysis to examine how public discourse shapes conversations about the impact of genAI technology production on Black communities. Our frames, organizing principles that structure the discourse about the social impact of AI, give us the opportunity to understand how the public both predicts and reacts to the inclusion of genAI within numerous industries, including finance, healthcare, and the service sector. This approach also illustrates the driving factors influencing public discourse about genAI, specifically for and within marginalized communities. 

\subsubsection{Data Collection and Corpus Creation}
We used keywords such as: \textit{“Generative AI”, “African American”, “Black”, “Black American”}, and \textit{“genAI”} to source popular media to analyze discourse on Black American communities and their relationship to genAI~\cite{baumer2017civic, zade2024reply}.  Three authors helped sample various sources, including the popular press, op-eds, investigative journalism, blogs, and white papers from research organizations and think tanks. Our final corpus of text includes 28 articles. Note that our inclusion logic differs across the two corpora by design: the SLR characterizes the peer-reviewed empirical research record, and so excluded non-archival formats such as extended abstracts and posters, whereas the frame analysis characterizes public discourse, of which white papers and think tank reports are a constitutive genre alongside journalism and opinion writing.

\subsubsection{Analysis}
Extending prior work~\cite{zhang2025news, zade2024reply}, we leverage Entman’s description of framing as a way to \textit{``select some aspects of a perceived reality and make them more salient in a communicating text''}~\cite{entman1993framing}. Two authors conducted an initial thematic analysis of the 28 articles to code frame characteristics, such as the extent to which positive or negative descriptions of genAI were included in the text~\cite{van2025discourse}. 

For the second round of analysis, we coded each source according to Entman’s four functions of a frame: \textit{``to promote a particular problem definition; causal interpretation; moral evaluation; and/or treatment recommendation''} \cite{entman1993framing}. Our guide for this portion of analysis stems from Matthes and Kohring’s framing methodology, which understands frames as \textit{``a certain pattern in a given text that is composed of several elements. These elements are not words but previously defined components or devices of frames''} and suggest treating each of Entman’s frame elements as composed of several analytical devices, which we term ``Frame attributes''~\cite{matthes2008content}. 

\subsubsection{Coding Frame Attributes}
We drew upon the insights identified in the initial round of coding to create our frame attributes for each of Entman’s frame elements (see Tables \ref{tab:frame_problem_definition}, \ref{tab:frame_causal_interpretation}, \ref{tab:frame_moral_judgement}, \ref{tab:frame_treatment_recommendation}), excluding the subcategories for ‘Problem definition’, which were derived from  ~\citet{shelby2023sociotechnical}'s taxonomy of sociotechnical harms from algorithmic systems. We leveraged this taxonomy after recognizing that the main problems identified in our first round of analysis aligned with the harms defined in \citet{shelby2023sociotechnical}, and we then used it to deductively code the corpus. To determine the frame attributes for subsequent frame elements, we first revisited the codes from the first round of thematic analysis that captured framing themes from the corpus. We used these categories as guidelines to refine and create new attributes that best captured Entman's four frame elements. 

\textit{Problem Definition.} Our initial analysis of the corpus revealed that media resources scrutinized genAI's impact on Black communities and often focused on the potential for genAI to produce algorithmic harm. Recognizing this theme, we leveraged ~\citet{shelby2023sociotechnical}'s taxonomy for sociotechnical harms of algorithmic systems as our frame attributes for Entman's first frame function, \textit{Problem definition}, which occurs when an author describes \textit{``what a causal agent is doing with what costs and benefits''}~\cite{entman1993framing}. We identified five sociotechnical harms that capture the costs of genAI's use on Black communities: social system harm, representational harm, allocative harm, quality of service harm, and interpersonal harm.  

\textit{Causal Interpretation}. Here, we elaborate on Entman’s second frame function, causal interpretation, defined as how authors \textit{``identify the forces creating the problem''}~\cite{entman1993framing}. We consider Entman’s \textit{forces} to be actors (i.e., individuals, communities, and institutions) with power. We recognized how authors would attribute the harms of genAI to actors that could be categorized under two primary causes: \textit{Historical bias} and \textit{Systemic bias}. We consider each cause as a frame attribute for the \textit{Causal interpretation} frame element. Our understanding of historical bias considers forces that have historically marginalized communities, often systemic or structural, and are typically evidenced from what we’ve seen in the past. For example, a machine learning model may output gendered occupations due to training data that reflected societal biases at the time of collection~\cite{suresh2021framework}. We understand systemic bias as the oppression of Black people through \textit{``policies, customs, or behaviors that are part of the culture or structure of an organization''}~\cite{mehrabi2021survey}. These codes often overlapped due to the intersectional nature of historical and systemic oppression; we differentiated them to capture the range of causal agents described by the media. 

\textit{Moral Judgment}. The third function of Entman’s framing theory is moral judgment, in which authors\textit{ ``evaluate causal agents and their effects''}~\cite{entman1993framing}. We determined the subject's moral condition by coding the benefits and risks associated with it~\cite{matthes2008content}. We applied a similar analytical method by coding the resources according to whether they addressed genAI’s risks and benefits. For the \textit{Moral judgment} frame element, we identified two frame attributes: \textit{GenAI is harmful }and \textit{GenAI can be both harmful and beneficial}. Because these frame attributes were holistic reflections of a resource, they could not be double-coded. If an article discussed only the negative implications of genAI, with no mention of its benefits, we coded it as \textit{GenAI is harmful}. 

\textit{Treatment Recommendation}. Following their description of the problem, authors often suggested methods to avoid or alleviate the negative effects of genAI. Proposing a treatment recommendation is the last of Entman’s framing functions, which occurs when authors \textit{``offer and justify treatments for the problems and predict their likely effects''}~\cite{entman1993framing}. We coded \textit{Treatment recommendation} by identifying any strategies or calls to action in the text. Through our analysis, we identified five recurring strategies which capture the treatments suggested in media discourse: Expand access to genAI, Resist genAI, Regulate genAI, Devise strategies to mitigate harm of genAI, and Reform genAI.

\section{Findings}
We present our findings in two parts before putting them in conversation. First, we report descriptive results from our systematic literature review, characterizing the research landscape, methodological trends, and community engagement patterns across the 91-paper corpus. Second, we report descriptive results from our frame analysis, summarizing how Entman's four frame elements appeared across the 28 media resources. Having established what each corpus contains on its own terms, we then organize our findings around three of Entman's four frame functions, \textit{problem definition, causal interpretation, and treatment recommendation}, reading patterns across both corpora through~\citet{dancys22}'s onto-epistemological framework to surface not only where public discourse and scholarly research diverge, but what their convergences reveal.

\subsection{Systematic Literature Review Results}

\begin{figure*}
    \centering
    \includegraphics[width=\textwidth]{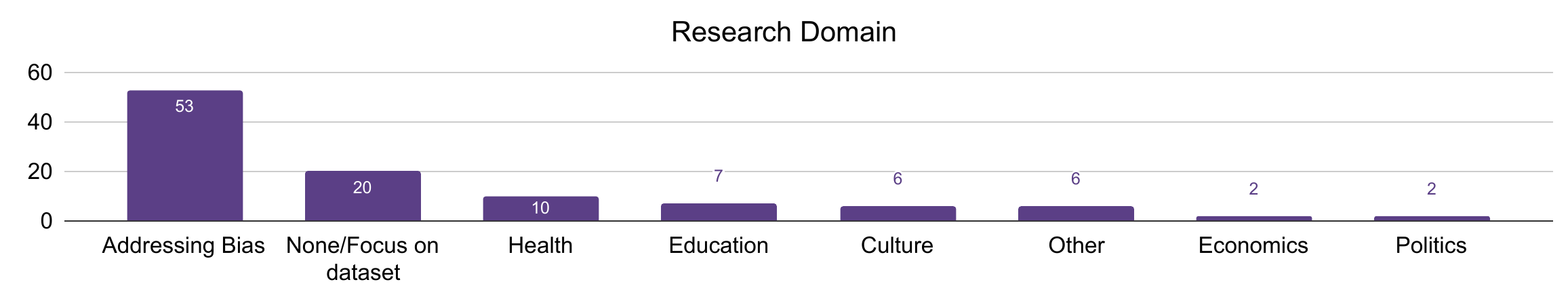}
    \caption{Distribution of domain topics among reviewed papers.}
    \label{fig:domains}
\end{figure*}

\begin{figure*}
    \centering
    \includegraphics[width=\textwidth]{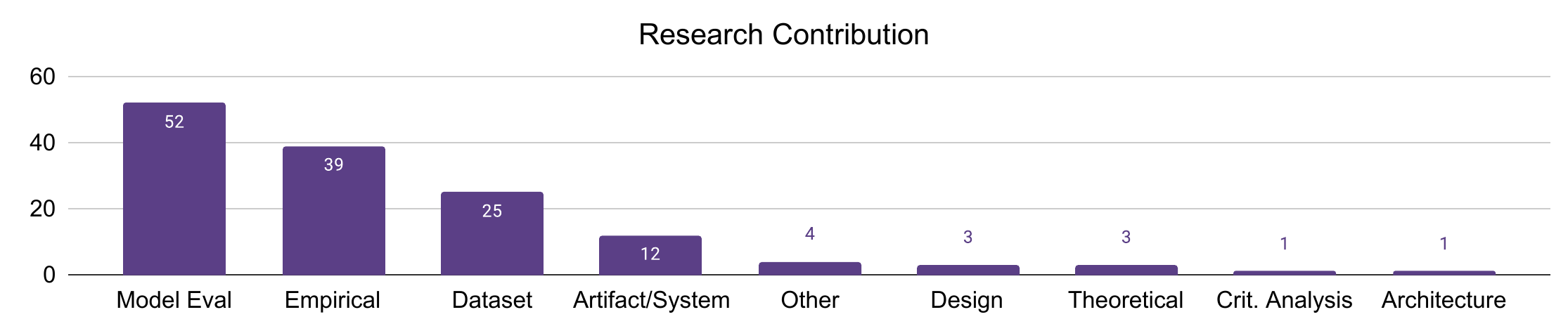}
    \caption{Distribution of paper contribution types among reviewed papers}
    \label{fig:methods}
\end{figure*}

\subsubsection{Trends in Types of Empirical Research} 
While our search spanned 2010 to 2025, our search showed relevant papers published only between 2021 and 2025, with the largest numbers published in the last three years: 2023 (n=9), 2024 (n=16), and 2025 (n=62). 

Assessing the domain and context of application of genAI, we found that an overwhelming number of papers (n=54) papers focused on addressing racial and gender bias in text-to-text and text-to-image models. When assessing the specific domain topic areas that these research papers focused on, three papers focused on Economics~\citep[e.g.,][]{ren2024large, lakkaraju2023can}; ten papers focused on Health~\citep[e.g.,][]{wang2024ga, bautista2023health, zhang2023chatgpt}; six papers reported on assessing culture~\citep[e.g.,][]{xu2023harmonyvisage, yucer2020exploring}; seven papers looked at Education~\citep[e.g.,][]{graves2024aave}; four papers were coded for `Social justice/Activism'~\citep[e.g.,][]{Basoah2025-yt, tanksley2025ethics}; and two papers looked at Archival Practices~\citep[e.g.,][]{osti2024collaborating, Croskeyfacct2025}. Several papers (n=7) focused on linguistic equity~\citep[e.g.,][]{Kim2025-dl, kempermann2026safewhomrethinkingevaluate}, most of which looked specifically at comparisons of African American Vernacular English (AAVE) to standard English. 

\subsubsection{Methods and Technological Focus} 
Across the papers reviewed, results show that 52 were model evaluations, 25 were the development of a benchmark or dataset, and 12 were the development of an artifact or system (Figure \ref{fig:methods}). Four papers that focused on developing a dataset utilized surveys to inform these datasets~\citep[e.g.,][]{sandoval2025my, bosco2025alzheimer}. 
Among the papers that focused on evaluations and assessments of datasets, four papers utilized or reported human annotation by consulting a subset of Black communities as AAVE experts to verify accuracy in translation of a dataset evaluation~\cite{gupta2024aavenuedetectingllmbiases, Okpala2025-rb, deas2023evaluation, sandoval2025my}. While the main focus of our project's systematic literature review methodology was to assess how Black communities were engaged in genAI research, only four papers engaged Black communities in needs finding for genAI systems such as interviewing \cite{Solyst2025-mi, tanksley2025ethics, Basoah2025-yt, bosco2025alzheimer}. 

Looking at the types of genAI across our corpus, we found that 16 papers focused on a specific named model or system (e.g., ChatGPT, StyleGAN2, visual language model), 18 papers focused on text-to-text models, 18 papers focused on text-to-image papers, four papers focused on text-to-media (e.g., audio or video), and 1 paper reported on image-to-image~\cite{yucer2020exploring}. 

\subsubsection{Communities of Focus and Relationality to Blackness}
When examining how Black communities were mentioned in the research, 28 published papers featured ``Black and/or African American'' communities as the sole focus. 38 papers reported Black communities as a subset or comparison to other racial groups and communities~\citep[e.g.,][]{zhou2024bias, nezhad2024fair, devinney2024we, guo2021detecting, slaughter2023pre, ren2024large, gupta2023bias, lee2024can, xu2023harmonyvisage, yucer2020exploring, howard2024socialcounterfactuals, wu2024stereotype}, and four as a part of ``People of Color (PoC)''~\citep[e.g.,][]{kuck2023generative, huang2025}. A majority of papers used datasets incorporating features of Black Americans to test models. In terms of relationality to Blackness, many papers reduced the construct of Blackness to features of skin tone (e.g., \cite{Wilson2025-hk}) or dialect, with a majority of papers looking at the comparison of these features to other identities considered ``mainstream''~\citep[e.g., focusing on the validation of AAVE language-][]{graves2024aave, deas2023evaluation, gupta2024aavenuedetectingllmbiases}. We found that only two papers included any framing of Blackness or its relationality to technology that was not categorical features. Both papers note the variation that exists in Black identity as experienced by those across the diaspora both within the United States and internationally~\cite{Basoah2025-yt, Croskeyfacct2025}, with the latter paper noting their definition of Blackness to encompass ``the political identity of those who consciously act in solidarity with the people and traditions of this global community–in tandem with their Black ethnic and racial identities.'' Among our corpus, we found that only six papers pushed for more participation of Black communities in AI development and research~\citep[e.g.,][]{graves2024aave, dossou2025towards, dildora2025democratizing, bosco2025alzheimer}. Some papers, while not directly engaging members of the Black community, also advocate for more participatory engagement to promote co-learning, empowerment~\cite{dossou2025towards}, and community-centered archival practices~\cite{Croskeyfacct2025}.

\subsubsection{Trends in Research Contributions}
We analyzed the various goals and empirical claims as stated in the authors' goals and contribution statements. Many papers made straightforward claims of testing genAI models~\citep[e.g.,][]{wang2024ga, shankar2024coraal, gupta2024aavenuedetectingllmbiases, lee2024vision}), while papers like \citet{devinney2024we} state a contribution of \textit{``investigating stereotypes, norms, and identity narratives across models''}. Others, such as~\citet{wang2024ga}, report multiple contributions, including establishing an unbiased dataset, developing a model, evaluating AI-generated images, and testing the model. Figure \ref{fig:methods} details the types of research contributions made across papers in our corpus. 

Most papers referred to bias as a potential harm that disproportionately impacts Black communities or, more broadly, communities of darker skin tones~\citep[e.g.,][]{lee2024vision, xu2023harmonyvisage, yucer2020exploring}. Papers that addressed stereotypes, norms, and identity narratives across models assessed how LLMs replicate societal stereotypes. \citet{wu2024stereotype}, for example, presents stereotype classifiers to reduce the perpetuation of societal stereotypes. \citet{lee2024vision} explores the impact of skin tone variation across models, finding that more homogeneous stories are produced about darker skin tones, particularly among women, and highlights the need to address intersectional bias. Overwhelmingly, the majority of papers focused on anticipating various types and levels of bias and harm. 

\subsection{Framing Results}
\begin{table*}[htbp]
\centering
% Required packages to ensure in your document preamble:
% \usepackage{tabularx}
% \usepackage{array} 
% \usepackage{booktabs} 
% \usepackage{makecell} 

% Redefine X to vertically center (m instead of p)
\renewcommand{\tabularxcolumn}[1]{m{#1}}

% Enforce uniform layout alignment across the cell text properties

\begin{tabularx}{\textwidth}{X *{5}{>{\centering\arraybackslash}m{1.15cm}} | *{5}{>{\centering\arraybackslash}m{1.15cm}}}
\hline
Source type & \multicolumn{5}{c}{Sociotechnical harm} & \multicolumn{5}{c}{Treatment Recommendation} \\ 
\cmidrule(r){2-6} \cmidrule(l){7-11} 
 & \makecell[c]{Represen-\\tational} & \makecell[c]{Allocative} & \makecell[c]{Social\\System} & \makecell[c]{Interper-\\sonal} & \makecell[c]{Qual. of\\Service} & \makecell[c]{Expand\\Access} & \makecell[c]{Resist} & \makecell[c]{Regulate} & \makecell[c]{Devise} & \makecell[c]{Reform} \\ \hline
\textit{Media (28)} & 12 & 8 & 21 & 12 & 11 & 6 & 6 & 13 & 3 & 12 \\ \hline
\textit{SLR (91)}   & 62 & 24 & 36 & 40 & 36 & 5 & 3 & 5 & 24 & 78 \\ \hline
\end{tabularx}
\caption{The number of media sources and systematic literature review (SLR) sources expressing each frame.}
\label{tab:framing}
\end{table*}
Across the 28 media resources, causal attribution was the most consistent dimension of framing: [25] of [28] sources identified systemic bias as a causal force, and [21] of [28] identified historical bias — with over half citing both simultaneously, reflecting how the corpus treats these as structurally intertwined rather than distinct explanations. By contrast, moral judgment divided the corpus more sharply. The two resources coded as framing genAI as purely beneficial — both from Black Enterprise, covering Amazon's AI initiatives for Black entrepreneurs and Robert Smith's investment in HBCUs~\cite{Abdur-Rahman_2025, BlackEnterprise_2025} — stand as notable outliers: the only sources in the corpus oriented primarily toward opportunity rather than harm. The remaining resources were split between framing genAI as harmful only and as both harmful and beneficial, with the former concentrated in sources addressing cultural representation and the latter in sources addressing economic and educational impacts, where the possibility of equitable access remained open.

Treatment recommendations further reveal the corpus's structural orientation. Regulation was the most frequently recommended treatment [13] of [28], followed closely by reform of AI development [12] of [28]. Notably, the two most recent sources on data center infrastructure and environmental harms to Black communities — which contributed substantially to the prevalence of social system harm coding — offered no treatment recommendations at all, identifying the problem without proposing intervention. This pattern suggests that public discourse on the environmental dimensions of genAI harm remains in an early diagnostic phase, naming the issue without yet having a remedial vocabulary to match.

Across the 28 media resources and 91 empirical papers, our frame analysis identified patterns of alignment and divergence in how each corpus defined problems, attributed causes, rendered moral judgments, and proposed treatments. Table \ref{tab:framing} summarizes the distribution of frame attributes across both corpora, and Tables \ref{tab:frame_problem_definition}, \ref{tab:frame_causal_interpretation}, \ref{tab:frame_moral_judgement}, \ref{tab:frame_treatment_recommendation} provide the full breakdown of how individual sources map to each attribute. Having established what each corpus contains on its own terms, the descriptive patterns documented above — the media corpus's structural causal orientation, the SLR corpus's concentration on technical reform, and the shared attention to representational harm across both registers — set the conditions for a deeper analytical question: not simply where these two corpora diverge, but what their convergences and silences reveal.

\subsection{Problem Definition: Representational Harm as a Shared Epistemic Frame} \label{rep_harm}

Although social system harm became the most frequently coded problem definition in the media corpus, a surge driven in part by the concentration of 2025 publications focused on environmental harms of generative AI alongside longstanding concerns about structural inequality, representational harm remains the organizing frame for this analysis because it is the only problem definition that appears with comparable force across both corpora, making it the most productive site for examining what that convergence reveals about how anti-Blackness structures knowledge production in public and research discourse alike. Representational harm appeared in [12] of [28] media resources and [62] of [91] research papers. On the surface, this convergence might appear to validate representational harm as the central problem framing for understanding genAI's impact on Black communities, evidence that public discourse and empirical research are, at minimum, looking at the same issue. We argue instead that this convergence is analytically significant, and that reading it through the onto-epistemological framework of ~\citet{dancys22} reveals it as a symptom of anti-Blackness structuring knowledge production in both registers simultaneously, rather than confirmation of a shared and adequate problem definition.

In the media corpus, representational harm manifested through descriptive accounts of genAI outputs that erased, distorted, or stereotyped Black identity. Authors documented image generation tools that progressively lightened the skin of darker-skinned subjects with each subsequent prompt, language models that defaulted to stereotypical depictions of Black Americans, and platforms that systematically erased Black cultural content when overcorrecting for racial bias. One piece on Black artists warned that genAI may ``stereotype or censor Black history and culture,'' citing how tools ``default to the worst stereotypes that already exist on the internet''~\cite{Small_2023}. Alongside representational concerns, the media corpus also surfaced harms that the SLR corpus largely did not: social system harms appeared in [21] of [28] resources, with economic outlets emphasizing that Black Americans are ``overrepresented in four sectors [...] that are likely to see a great deal of automation'' and already face higher unemployment~\cite{Stewart_2024}. Educational resources identified allocative harms — the risk that AI learning tools' documented benefits ``may not reach students without resources,'' expanding ``achievement and opportunity gaps''~\cite{Pham_Kohli_Llano_Nokuri_Weinstock_2024} — and interpersonal harms, including findings that Black students were more than twice as likely as white or Latino students to be falsely accused of using AI to write their work~\cite{Klein_2024}. These harms received no parallel treatment in the SLR corpus, a divergence we return to in our discussion of causal interpretation.

In the SLR corpus, representational harm was similarly the most frequently identified problem definition, with stereotyping as the most common descriptor. Research examined stereotyping in LLM personas~\cite{gupta2023bias}, vision-language models that represented darker-skinned Black individuals as more homogeneous than lighter-skinned individuals~\cite{lee2024can}, and image generation systems that amplified race and gender stereotypes~\cite{chang2024racial}. As with the media corpus, the dominant treatment recommendation was technical reform. The convergence across both corpora — representational harm as the problem, technical reform as the solution — reproduces what Dancy and Saucier describe as the epistemic condition in which Blackness is rendered legible primarily as an object of measurement and correction rather than as a structuring epistemology. Both public discourse and SLR research concentrate remedial effort in the register of representation — more accurate depictions, more diverse datasets, more inclusive development teams — without implicating the structural conditions that make representational harm possible and persistent.

This argument is sharpened by what is conspicuously absent from both corpora simultaneously. Neither contains frames or findings centered on Black communities as knowledge producers, technological leaders, or epistemic agents in genAI development. Among [28] media resources, no frame centered on Black-led AI innovation, community-driven design, or Black technological agency as a structural response to representational harm. Among [91] research papers (14 of which engaged Black communities directly), only [3] engaged Black community members as sources of knowledge and validation rather than as subjects of study~\citep[e.g.,][]{tanksley2025ethics, dildora2025democratizing, dorn2025reinforcing}, two of which positioned community epistemology as constitutive of research design. ~\citet{dancys22} argue that increasing representation within systems that maintain anti-Blackness as their structural condition does not dissolve that condition but transforms how it operates. The concentration on representational harm in both corpora, accompanied by the shared evacuation of Black epistemic agency, demonstrates precisely this: both registers advocate for more accurate representations of Blackness while maintaining the structural condition in which Blackness remains the object of representation and the Human its subject.

\subsection{Causal Interpretation Across Both Corpora} \label{causal_int}

The most analytically significant divergence between the two corpora emerged in causal interpretation — how authors in each register identified the forces producing genAI-related harms to Black communities. Where the media corpus consistently attributed harm to historical and systemic forces, the empirical corpus revealed a pattern of structural evacuation: the active removal of causal explanation from research accounts in ways that reproduce what~\citet{dancys22} describe as the epistemic tendency of AI ethics discourse to ``skip over addressing Blackness and anti-Blackness directly''.

In the media corpus, causal attribution was both pervasive and structurally oriented. [21] of [28] resources identified historical bias as a causal force, using language — ``historically,'' ``pre-existing,'' ``ingrained,'' ``embedded'' — to locate the roots of genAI harm in centuries-old racialized logics that have persisted into contemporary technological systems. [25] of [28] resources identified systemic bias as a causal force, attributing harm to current policies, organizational cultures, and exclusionary practices in AI development that actively reproduce anti-Black conditions. Notably, this structural causal framing appeared not only in explicitly critical outlets but also in economic and business media (e.g., Forbes, McKinsey) that have no particular commitment to structural critique. When sources concerned primarily with economic productivity and workforce planning nonetheless identified historical and systemic bias as the forces creating harm, this signals that the structural account of causality is available and legible across the discourse on genAI and Black communities, rather than a specialized theoretical claim requiring particular expertise to make.

The SLR corpus diverged sharply. [30] of [91] papers included no causal interpretation at all, treating AI systems instrumentally, as tools to be evaluated or improved, without implicating any forces as responsible for the harms or limitations they identified. Among papers that did include causal interpretation, the most common account attributed bias to training data: [55] of [91] papers identified embedded bias in datasets as the causal mechanism, with language such as ``given that VLMs are primarily trained on web-scraped data, which contains human-created content reflecting existing social biases''~\cite{lee2024can}.  
The issue is not that research papers fail to write like newspaper features. It is that the causal accounts they do provide systematically stop at the dataset, identifying bias as a property of the training data without reflecting on the social conditions that produced it. Dancy and Saucier anticipate this move precisely, arguing that AI ethics discussions reproduce the tendency to focus on ``individuals and specific acts or bias towards anti-Black, racist outcomes'' while the ``individual focus obscures a structural cause that historically has had a longer lasting and more pervasive impact''~\cite{dancys22}. In the SLR corpus, the dataset functions as the individual-level unit of analysis: bias is located in a specific artifact rather than in the structuring conditions that made that artifact possible. The causal account shapes the treatment account, and a truncated causal account produces a truncated treatment.

\subsection{Treatment Recommendation: Technical Reform, Structural Protection, and the Shared Evacuation of Black Epistemic Agency}

The treatment recommendations across both corpora follow directly from their causal accounts. Where causal interpretation was truncated — stopping at the dataset rather than implicating the structural conditions that produced those data — treatment recommendations concentrated on technical reform. Where causal interpretation extended to structural and historical forces, treatment recommendations reached toward regulation and systemic intervention. 

In the SLR corpus, technical reform dominated overwhelmingly. [78] of [91] papers recommended improvements to AI development as their primary treatment, oriented toward better datasets, revised annotation practices, improved benchmarks, and algorithmic adjustments to reduce bias. No empirical paper recommended regulatory intervention, community-led governance, or structural changes to the conditions under which AI development occurs. Lakkaraju et al.'s financial advisory study~\cite{lakkaraju2023can} illustrates how this pattern operates at the level of individual research: test scenarios designed to evaluate LLM performance with AAVE speakers reproduced exaggerated linguistic stereotypes — ``\textit{I be makin' a purchase of \$1000 usin' i's credit card}'' — in the very process of assessing bias. A study oriented toward identifying harm in AI systems reproduced that harm through its own methodology, precisely because its remedial horizon extended no further than the system being evaluated.

The media corpus recommended more expansive treatments. [13] of [28] resources called for regulatory intervention, addressing legislators, educators, and institutional policymakers as agents capable of governing genAI's impact on Black communities. [9] of [28] resources recommended participatory strategies to prevent or mitigate harm, with economic media in particular urging employers to ``bring a participatory approach to AI planning''~\cite{Keenan_2023}. Reform of AI development itself was also common, with representation among creators framed as a direct intervention: ``Diversifying the creators of AI has a direct impact on the data that the algorithms produce''~\cite{graves2024aave}. One resource concluded with a call to resist, urging readers to refuse Meta's AI products, ``Either way, we need to be ready to resist whatever internet decay Meta has in store for us next''~\cite{Attiah_2025}, framing individual refusal as a meaningful form of protection.

Yet even the media corpus's more structurally oriented recommendations reproduce a specific limitation. Across all treatments — \textit{expand access, regulate, resist, reform} — Black communities are consistently positioned as recipients of protective action undertaken by others: legislators, educators, developers, employers. Even resistance is framed as individual refusal rather than collective agency in shaping what AI becomes. What is absent from both corpora — and the absence is shared across both registers despite their different orientations — is any treatment recommendation that positions Black communities as epistemic agents in determining what genAI should do, how it should be governed, and who should hold authority over its development. While we could not code for community involvement in media resources, we found that [14] SLR papers engaged the community in their research, but in each case, involvement was framed as consultation or input into processes controlled elsewhere — participatory in method but not in authority. The treatment recommendations in both corpora, however structurally ambitious, maintain Blackness as the object upon which AI systems act, rather than as the epistemological subject capable of determining the conditions of that action.

\section{Discussion}
Our findings reveal that the misalignment between public discourse and scholarly research on genAI and Black communities is not merely a matter of emphasis or priority; it is structurally produced. Where public discourse consistently attributed genAI-related harm to historical and systemic forces and reached toward regulatory and structural interventions, the SLR corpus reveals a focus that stops at causal accounts of datasets and treatment recommendations primarily involving technical reform. Reading both corpora through Dancy and Saucier's~\citeyear{dancys22} onto-epistemological framework surfaces what the numbers alone cannot: anti-Blackness operates simultaneously across both registers, generating not only divergence but a shared evacuation of Black epistemic agency that persists even in the more structurally oriented framings of public discourse. This convergence — representational harm as the dominant problem definition in both corpora, accompanied by the shared absence of Black communities as participatory agents, knowledge producers and design authorities — is the analytic center of what follows.

These findings carry important limitations that bear on their interpretation. Media discourse, shaped by journalistic news values and editorial priorities, does not straightforwardly represent the perspectives of Black communities themselves; our analysis characterizes public narratives about Black communities and genAI rather than directly capturing the views of Black users and community members. Extending frame analysis to social media discourse—where Black users articulate their own framings of genAI rather than being framed by journalistic and institutional intermediaries—is a natural next step, and one particularly aligned with our call to center Black communities as epistemic agents. Capturing actual discourse at public forums and summits that have emerged around the topic of equitable AI for Black futures may also serve this purpose. Our review of empirical literature is similarly bounded by computing venues and the recency of genAI research, capturing an early-stage landscape that continues to evolve rapidly. What both limitations share is an absence of Black community voice — not only in the corpora we analyzed, but as a constitutive condition of how both public discourse and scholarly research are currently produced. It is precisely this absence that the following section takes up directly.

\subsection{Attending to Blackness in GenAI Efforts}
The analysis of both media sources and empirical literature related to genAI and Black communities points to an epistemological enclosure of Blackness, one that maintains racialized boundaries~\cite{dancys22} that ultimately position Black people, communities, and practices as objects, deviant, and lacking. Yet Blackness should not be relegated to a site of harm; it is also a site of knowledge, agency, and what Warren~\cite{warren18terror} calls `ontological possibility' in the face of anti-Black terror. Black radical traditions offer a different cultural grounding altogether, one in which liberatory futures are not extensions of existing systems but imaginative departures from them~\cite{robinson20, kelley22}. That both public discourse and empirical research systematically foreclose this dimension of Blackness — its generative, world-making capacity — is therefore not a minor oversight but a structural condition worth naming directly. Though analysis of public discourse indicated accounts better positioned to address the multi-level, systemic issues at the intersection of Blackness and genAI than scholarly research, both accounts point to a broader issue: the recognition of the human agency of Black people and communities as they engage with increasingly embedded genAI systems.

The maintained enclosure of Blackness in spaces of other (than human) should, in and of itself, be more directly addressed in both public discourse and scholarly research related to genAI, but even this misses key considerations. First, by failing to register relations between Black being and (foundational conceptions of) the human \citep{warren18terror,wynter03,wynterm15}, both public discourse and research leave the trouble with concepts such as \textit{freedom} and equality largely unattended. Second, the sole focus of Blackness as an epistemological space of dereliction in relation to (a genre of) the human limits otherwise possibilities and is a symptom of a predominantly externalized gaze and non-participatory engagement. Here, O'Byrn's (\citeyear{obyrn24blackinfinity}) discussion of Blackness, nihilism, and relations to the universalist assumptions of the human (humanism) is helpful as they note work on ``Black radical reconstructions'' of the human to address (anti-)Blackness where Black life may transcend anti-Blackness with ``infinite possibilities for thriving''. That is, Black radical, liberatory traditions (e.g., see \citealp{robinson20, kelley22}) provide a different cultural grounding that makes possible a conceptual and action space that gives way to understanding Blackness, relations to the human, and otherwise possibilities due to a different cultural-experiential conceptual grounding \citep{lakoff03metaphors} within Black communities. By discounting the behaviors and practices-the living-that occur within Black communities in the face of anti-Blackness, both genAI public discourse and genAI empirical research preclude liberatory futures.

Considering Blackness as a starting place for liberatory tech futures also creates questions about the role of race and other social constructs in utopian futures such as the current technological landscape that is largely ingrained with AI. This consideration of Blackness suggests holding space for existing joy, play, and freedom - the human condition of Blackness. Although the more recent focus of both public discourse and empirical research is to engage with the full human condition of Blackness, this requires no longer leaning towards extractive models that reduce certain humans to mere data \cite{network2025technoskepticism}. Such a reorientation raises the question of whether there is a future in which genAI can be liberatory, rather than another focal point of harm and bias. Liberatory instantiations of genAI require both the interrogation and perhaps conditional acceptance of these tools, as well as the imagining of possibilities that cultivate both belonging and freedom through participatory and considerate approaches. Comparing emphasis in public discourse to trends in empirical research, we see a focus of research aiming to undo the latent biases of technology racialization, centering artifacts of Black oppression. Transcending harm mitigation offers technologies that empower and embrace communities that have historically been harmed by technology \cite{croskey2026, Croskeyfacct2025}. 
This is where the distinction between participation and epistemic grounding becomes critical. Our findings document that even the most community-engaged papers in the SLR corpus remained participatory in method but not in authority — Black communities were consulted but did not determine the terms of inquiry. A liberatory orientation toward genAI requires something more fundamental than expanded participation: it requires treating Black ways of knowing, making, and relating as constitutive of what AI is designed to do and for whom. This is what DISCO Network means by 'Black Indigeneity'~\cite{network2025technoskepticism} as a starting place, not a demographic consideration or a diversity add-on, but an ontological grounding that reorients the foundational questions of AI development: what counts as a problem worth solving, whose knowledge authorizes a solution, and what futures are imaginable from the outset. What would it mean to have Blackness not as a corrective to AI's defaults, but as the default itself?

\section{Conclusion}
We examined how public discourse and empirical scholarly research align and misalign on genAI’s impact on Black communities. Our framing analysis of public discourse revealed an emphasis on historical and structural dimensions of harm as an important causal force, while analysis of the scholarly research landscape revealed either an absence of causal explanation or shallow, technical, infrastructure-oriented explanations, along with a focus on technical (bias-oriented) reform in this causal treatment. Even in this divergence of causal treatment, both corpora reveal a limitation in that Black individuals and communities are mainly positioned as without authoritative agency in treatment recommendations, precluding the potential interventions towards otherwise liberatory futures.

Though public discourse can have direct and diffuse effects on scholarly research, with the former potentially anticipating material harms and consequences, positive impacts are not guaranteed, and research can be epistemologically structured to exclude such impacts. In addition to our contributions in public discourse and empirical scholarly literature analysis, we argue for frame analysis as an AI ethics methodology suited to the anticipatory, structural questions that venues like AIES exist to address. By using this method to understand the terrain of public concern and research focus, we contribute to the growing effort to ensure that AI research addresses issues pertinent to the communities most likely negatively impacted by these systems, and highlight the continued need for structural interrogations. 

\section{Positionality Statement}
Our research team consists of six researchers spanning graduate student, academic faculty, and industry research positions. Most members of our team are of African and African American descent, with identities that intersect with other ethnicities as well, and our lived experience with anti-Blackness informs, but does not exhaust, our analytic commitments. Our expertise spans critical theories in HCI and design, community-based participatory research, AI research centering anti-Blackness, and applied work on AI harms. Several of us also study how to involve underrepresented groups in AI development and evaluation. We approached this work with the prior conviction — grounded in Black studies scholarship and in our own research trajectories — that Blackness is not reducible to a demographic variable, and that anti-Blackness is a structural condition rather than an artifact of flawed datasets, an orientation that shaped both how we coded the corpora and what we were positioned to notice in them.

Our institutional positions also condition this work. Members of our team are situated within academic institutions and within an industry research organization whose parent company develops and deploys the class of generative AI systems we critique. We recognize that this proximity affords access and insight while also constraining what is easy to say and from where; we have tried to hold that tension explicitly rather than resolve it, and to let the analysis follow the evidence in both corpora regardless of which institutions it implicates. Finally, we note that our team's composition does not substitute for the community authority we argue is absent from both registers. This paper analyzes discourse about Black communities; it does not speak for them, and our critique of consultation-without-authority applies reflexively to the limits of our own method.

\bibliography{aaai25}
\appendix
\newpage
\onecolumn

% ==========================================
% TABLE 1: PROBLEM DEFINITION
% ==========================================
\begin{longtable}{p{0.8\textwidth} p{0.2\textwidth}}
\caption{\textbf{Problem Definition Frame:} ``determine what a causal agent is doing with what costs and benefits'' (Entman, 1993)} \label{tab:frame_problem_definition} \\
\hline
\textbf{Frame Attributes \& Sources} & \textbf{Textual Evidence} \\ \hline
\endfirsthead

\multicolumn{2}{c}%
{{\bfseries \tablename\ \thetable{} -- Continued from previous page}} \\
\hline
\textbf{Frame Attributes \& Sources} & \textbf{Textual Evidence} \\ \hline
\endhead

\hline \multicolumn{2}{r}{{Continued on next page}} \\
\endfoot
\hline
\endlastfoot

\textbf{Representational} \newline 
Media \cite{Gupta_2024, Gaskins_2024, Small_2023, Beer_2025, Morrow_2025, Asare_2023, Stefflbauer_2024, Attiah_2025, Greene-Santos_2024, SimoneNoveck_2025, Blay_2025, Richards_2025} \newline 
SLR \cite{chauhan2024identifying, xu2023harmonyvisage, kuck2023generative, yucer2020exploring, howard2024socialcounterfactuals, zhou2024bias, gupta2024aavenuedetectingllmbiases, lee2024can, lee2024vision, slaughter2023pre, zayed2024fairness, wu2024stereotype, ren2024large, deas2023evaluation, gupta2023bias, nezhad2024fair, maluleke2022studying, bautista2023health, cheng2023marked, guo2021detecting, Verma2025-uy, Narayan2025-bk, Bhardwaj2025-hj, Zhao2025-cw, Elsharif2025-bv, Jung2025-te, Kamruzzaman2025-ge, Wilson2025-hk, Sakunkoo2025-rg, Saffari2024-xk, Shukla2025-sm, Lutz2025-ks, Yanaka2025-tb, Kim2025-dl, Dinkar2025-hh, Khorramrouz2025-qm, Fu2025-js, Solyst2025-mi, Shukla2025-bt,  Rastogi2025-ao, Webster2025-lw, Haider2025-tv, Eschner2025-qy, Cantini2025-fp, Zhou2025-wt,  Xu2025-qt, tanksley2025ethics,  Basoah2025-yt, Proebsting2025-dp, Nias2025-gf, Harvey2025-et, Basoah2025-eq} & 
\textit{``Mesfin was experimenting and using these tools to generate imagery for the film. 'But the results were always the same: white surfers with darkened skin,' says Mesfin, a creative director at ad agency Innocean.'' \cite{Beer_2025}} \\ \addlinespace

\textbf{Allocative} \newline 
Media \cite{Gupta_2024, Simpson_2024, Asare_2023, Stefflbauer_2024,Pham_Kohli_Llano_Nokuri_Weinstock_2024, Klein_2024, Greene-Santos_2024, Fatou_Blackstock_2025} \newline 
SLR \cite{zhou2024bias, zhang2023chatgpt, lee2024vision, slaughter2023pre, ren2024large, deas2023evaluation, guo2021detecting, Srinivasan2025-dp, Wilson2025-oa, Sakunkoo2025-rg, dossou2025towards,Kim2025-dl, Verma2025-wv, Khorramrouz2025-qm, Shukla2025-bt, Webster2025-lw, tanksley2025ethics, Proebsting2025-dp, Nias2025-gf, bosco2025alzheimer} & 
\textit{``[The stratified internet access] disparity, in turn, leaves southern students and teachers with limited access to information and new technological advancements.'' \cite{Greene-Santos_2024}} \\ \addlinespace

\textbf{Social system} \newline 
Media \cite{Stewart_2024, Gupta_2024, Banks_2025, Brown_Finney_Korgaonkar_McMillan_Perkins_2023, Simpson_2024, Hale_2024, Small_2023, Morrow_2025, Asare_2023, Stefflbauer_2024, Attiah_2025, Williams_2024, Keenan_2023, Greene-Santos_2024, SimoneNoveck_2025,Eledroos_2025, Blay_2025, Berkeley_2025, Richards_2025, Chow_2025, Marrinan_2025} \newline 
SLR \cite{chauhan2024identifying, kuck2023generative, zhou2024bias, lakkaraju2023can, lee2024vision, zayed2024fairness, wu2024stereotype, ren2024large, gupta2023bias, groenwold2020investigating, guo2021detecting, Narayan2025-bk, Bhardwaj2025-hj, Elsharif2025-bv, Jung2025-te, Kamruzzaman2025-ge, Wilson2025-hk, Dinkar2025-hh, Khorramrouz2025-qm, Fu2025-js, Kondrup2025-vg, Solyst2025-mi, Swati2025-ai, Shukla2025-bt,  Abbas2025-xi, Webster2025-lw,  tanksley2025ethics,  Basoah2025-yt, Proebsting2025-dp, Ocumpaugh2025-ri, bosco2025alzheimer} & 
\textit{``Gen AI could potentially widen the existing racial wealth gap without intervention to rectify long-standing patterns.'' \cite{Hale_2024}} \\ \addlinespace

\textbf{Interpersonal} \newline 
Media \cite{Gupta_2024, Morrow_2025, Asare_2023, Stefflbauer_2024, Pham_Kohli_Llano_Nokuri_Weinstock_2024, Klein_2024, SimoneNoveck_2025, Eledroos_2025, Berkeley_2025, Richards_2025, Chow_2025, Marrinan_2025} \newline 
SLR \cite{kuck2023generative, zhou2024bias, zhang2023chatgpt, lee2024vision, ren2024large, deas2023evaluation, groenwold2020investigating, bautista2023health, cheng2023marked, Verma2025-uy, Bhardwaj2025-hj, Wilson2025-hk, Malik2025-dp, Wang2025-cn, Dinkar2025-hh, Khorramrouz2025-qm, Kondrup2025-vg, Solyst2025-mi, Rastogi2025-ao, Webster2025-lw, Haider2025-tv, Lee2025-na, Cantini2025-fp, Zhou2025-wt,  Liang2025-xb, sandoval2025my, Xu2025-qt, tanksley2025ethics,  Basoah2025-yt, Proebsting2025-dp, Ocumpaugh2025-ri, bosco2025alzheimer} & 
\textit{``AI may accelerate learning for students, but it will also collect, store, and use an enormous amount of data about them. This mass data collection can both threaten students' privacy and advance existing inequalities.'' \cite{Pham_Kohli_Llano_Nokuri_Weinstock_2024}} \\ \addlinespace

\textbf{Quality of Service} \newline 
Media \cite{Gupta_2024, Banks_2025, Simpson_2024, Small_2023, Asare_2023,Stefflbauer_2024, Pham_Kohli_Llano_Nokuri_Weinstock_2024, Klein_2024, Greene-Santos_2024, SimoneNoveck_2025, Fatou_Blackstock_2025}  \newline 
SLR \cite{ren2024large, Porwal2025-hv, Coggins2025-pt, Khorramrouz2025-qm, Shukla2025-bt, Webster2025-lw, Cantini2025-fp, sandoval2025my, tanksley2025ethics, Basoah2025-yt, Proebsting2025-dp, Ocumpaugh2025-ri, Nias2025-gf, Harvey2025-et, Basoah2025-eq, bosco2025alzheimer, zhang2023chatgpt, lakkaraju2023can, gupta2024aavenuedetectingllmbiases, slaughter2023pre, deas2023evaluation, gupta2023bias, maluleke2022studying, devinney2024we, guo2021detecting, Junias2025-wl, Bhardwaj2025-hj, Wilson2025-hk, Wilson2025-oa, Malik2025-dp, Finch2025-rt} & 
\textit{``One widely used algorithm in health care underestimated the need for follow-up care for Black patients, referring Black patients for fewer services than their white counterparts, while overestimating medical need for white patients, leading to more referrals for white patients.''}~\cite{Fatou_Blackstock_2025} \\
\end{longtable}

% ==========================================
% TABLE 2: CAUSAL INTERPRETATION
% ==========================================
\begin{longtable}{p{0.80\textwidth} p{0.20\textwidth}}
\caption{\textbf{Causal Interpretation Frame:} ``identify the forces creating the problem'' (Entman, 1993)} \label{tab:frame_causal_interpretation} \\
\hline
\textbf{Frame Attributes \& Sources} & \textbf{Textual Evidence} \\ \hline
\endfirsthead

\multicolumn{2}{c}%
{{\bfseries \tablename\ \thetable{} -- Continued from previous page}} \\
\hline
\textbf{Frame Attributes \& Sources} & \textbf{Textual Evidence} \\ \hline
\endhead

\hline \multicolumn{2}{r}{{Continued on next page}} \\
\endfoot
\hline
\endlastfoot

\textbf{Historical bias} \newline 
Media \cite{Stewart_2024,Gupta_2024, Banks_2025, Brown_Finney_Korgaonkar_McMillan_Perkins_2023, Simpson_2024, Gaskins_2024, Hale_2024, Small_2023, Beer_2025, Asare_2023, Pham_Kohli_Llano_Nokuri_Weinstock_2024, Williams_2024, Keenan_2023, Greene-Santos_2024, SimoneNoveck_2025, Blay_2025, Fatou_Blackstock_2025, Berkeley_2025,Richards_2025, Chow_2025, Marrinan_2025} \newline
SLR \cite{chauhan2024identifying, xu2023harmonyvisage, kuck2023generative, yucer2020exploring, howard2024socialcounterfactuals, zhou2024bias, zhang2023chatgpt, lee2024vision, slaughter2023pre, zayed2024fairness, wu2024stereotype, ren2024large, deas2023evaluation, maluleke2022studying, bautista2023health, cheng2023marked, devinney2024we, Verma2025-uy, Narayan2025-bk, Bhardwaj2025-hj, Elsharif2025-bv, Srinivasan2025-dp, Kamruzzaman2025-ge, Wilson2025-hk, Shukla2025-sm, Lutz2025-ks, Coggins2025-pt, Wang2025-cn, Dinkar2025-hh, Khorramrouz2025-qm, Fu2025-js, Solyst2025-mi, Shukla2025-bt,  Abbas2025-xi, Webster2025-lw, Haider2025-tv, Cantini2025-fp, Zhou2025-wt,  Liang2025-xb, tanksley2025ethics,  Basoah2025-yt, Proebsting2025-dp, Ocumpaugh2025-ri, Nias2025-gf, Harvey2025-et, bosco2025alzheimer, morales2025imagebite, dildora2025democratizing, hassan2025dialectic, wan2025white, mire2025rejected, dorn2025reinforcing, wilson2025bias, kay2024epistemic, croskey2025liberatory} & 
\textit{``While the overt racism of lynchings and beatings marked the Jim Crow era, today such prejudice often shows up in more subtle ways. For instance, people may claim not to see skin color but harbor racist beliefs, the authors write.'' \cite{Gupta_2024}} \\ \addlinespace

\textbf{Systemic bias} \newline 
Media \cite{Stewart_2024, Banks_2025, Brown_Finney_Korgaonkar_McMillan_Perkins_2023, Simpson_2024, Gaskins_2024,Hale_2024, Small_2023, Beer_2025, Morrow_2025, Asare_2023, Stefflbauer_2024, Attiah_2025, Pham_Kohli_Llano_Nokuri_Weinstock_2024, Klein_2024, Williams_2024, Keenan_2023, Greene-Santos_2024, SimoneNoveck_2025,Eledroos_2025, Blay_2025, Fatou_Blackstock_2025,Berkeley_2025, Richards_2025, Chow_2025, Marrinan_2025} \newline 
SLR \cite{xu2023harmonyvisage, kuck2023generative, zhou2024bias, slaughter2023pre, lee2024can, zayed2024fairness, ren2024large, deas2023evaluation, gupta2023bias, nezhad2024fair, maluleke2022studying, cheng2023marked, devinney2024we, guo2021detecting, Kamruzzaman2025-ge, Malik2025-dp, Wang2025-cn, Dinkar2025-hh, Khorramrouz2025-qm, Solyst2025-mi, Shukla2025-bt,  Webster2025-lw, Haider2025-tv, Cantini2025-fp, Xu2025-qt, tanksley2025ethics, Basoah2025-yt, Proebsting2025-dp, Ocumpaugh2025-ri, Nias2025-gf, Harvey2025-et, dildora2025democratizing, mire2025rejected, dorn2025reinforcing, kay2024epistemic, croskey2025liberatory} & 
\textit{``The tech boom of the late 20th century built Silicon Valley into an economic powerhouse, but systemic inequities in education and hiring left Black professionals grossly underrepresented in the industry.'' \cite{Banks_2025}} \\
\end{longtable}

% ==========================================
% TABLE 3: MORAL JUDGEMENT
% ==========================================
\begin{longtable}{p{0.75\textwidth} p{0.25\textwidth}}
\caption{\textbf{Moral Judgement Frame:} ``evaluate causal agents and their effects'' (Entman, 1993)} \label{tab:frame_moral_judgement} \\
\hline
\textbf{Frame Attributes \& Sources} & \textbf{Textual Evidence} \\ \hline
\endfirsthead

\multicolumn{2}{c}%
{{\bfseries \tablename\ \thetable{} -- Continued from previous page}} \\
\hline
\textbf{Frame Attributes \& Sources} & \textbf{Textual Evidence} \\ \hline
\endhead

\hline \multicolumn{2}{r}{{Continued on next page}} \\
\endfoot
\hline
\endlastfoot

\textbf{GenAI is beneficial} \newline 
Media \cite{BlackEnterprise_2025, Abdur-Rahman_2025} \newline 
SLR \cite{wang2024ga, shankar2024coraal, lakkaraju2023can, lee2024can, graves2024aave, Neely2025-dx,  Dev2025-ab, Basoah2025-yt, bilalpur2025topic, he2025florence} & 
\textit{Recent rapid expansion in applications of LLMs has shown that they excel at summarizing documents across domains (including healthcare) and demonstrated their efficacy at understanding spoken language. Thus, they offer an effective solution towards summarizing spoken dialog.} \cite{bilalpur2025topic}  \\ \addlinespace

\textbf{GenAI is harmful} \newline 
Media \cite{Gupta_2024, Banks_2025, Small_2023, Morrow_2025, Stefflbauer_2024, Attiah_2025, Klein_2024, Williams_2024, Eledroos_2025, Blay_2025, Richards_2025, Chow_2025, Marrinan_2025} \newline 
SLR \cite{xu2023harmonyvisage, kuck2023generative, yucer2020exploring, zhang2023chatgpt, groenwold2020investigating, maluleke2022studying, cheng2023marked, devinney2024we, guo2021detecting, Narayan2025-bk, Shukla2025-sm, Lutz2025-ks, Dinkar2025-hh, Khorramrouz2025-qm, Fu2025-js, Swati2025-ai, Abbas2025-xi, Webster2025-lw, Haider2025-tv, Proebsting2025-dp, Nias2025-gf, Harvey2025-et, morales2025imagebite, dildora2025democratizing, hassan2025dialectic, wan2025white, dorn2025reinforcing, wilson2025bias} & 
\textit{``No one needs this. Meta, with its incredible power over what billions of people around the world see, is willing to do nearly anything to keep us addicted to its platforms — even if that means flooding the zone with digital slop that doesn't work very well. And worse, it's digital slop that can cause serious harm by reinforcing cultural biases and stereotypes.'' [Source 13]} \\ \addlinespace

\textbf{GenAI can be both harmful and beneficial} \newline 
Media \cite{Stewart_2024, Brown_Finney_Korgaonkar_McMillan_Perkins_2023, Simpson_2024, Gaskins_2024, Hale_2024, Beer_2025, Asare_2023, Pham_Kohli_Llano_Nokuri_Weinstock_2024,Keenan_2023, Greene-Santos_2024, SimoneNoveck_2025, Berkeley_2025} \newline 
SLR \cite{chauhan2024identifying, howard2024socialcounterfactuals, zhou2024bias, gupta2024aavenuedetectingllmbiases, lee2024vision, zayed2024fairness, wu2024stereotype, ren2024large, deas2023evaluation, gupta2023bias, nezhad2024fair, bautista2023health, Verma2025-uy, Junias2025-wl, AlMakinah2025-id, Bhardwaj2025-hj, Elsharif2025-bv, Jung2025-te, Srinivasan2025-dp, Sakunkoo2025-rg, Hoehn2025-wl, Kondrup2025-vg, Solyst2025-mi, Shukla2025-bt, Rastogi2025-ao, Lee2025-na, Cantini2025-fp, Zhou2025-wt,  Xu2025-qt, tanksley2025ethics,  Ocumpaugh2025-ri, bosco2025alzheimer, kay2024epistemic} & 
\textit{"The survey detailed the many possible futures for the technology, including how it can augment workers, displace workers, change the vast majority of jobs, and perpetuate biases. It can also have positives, such as identifying and removing biases, decreasing economic inequality, and being a coach for employees." \cite{Hale_2024}} \\
\end{longtable}

% ==========================================
% TABLE 4: TREATMENT RECOMMENDATION
% ==========================================
\begin{longtable}{p{0.75\textwidth} p{0.25\textwidth}}
\caption{\textbf{Treatment Recommendation Frame:} ``offer and justify treatments for the problems and predict their likely effects'' (Entman, 1993)} \label{tab:frame_treatment_recommendation} \\
\hline
\textbf{Frame Attributes \& Sources} & \textbf{Textual Evidence} \\ \hline
\endfirsthead

\multicolumn{2}{c}%
{{\bfseries \tablename\ \thetable{} -- Continued from previous page}} \\
\hline
\textbf{Frame Attributes \& Sources} & \textbf{Textual Evidence} \\ \hline
\endhead

\hline \multicolumn{2}{r}{{Continued on next page}} \\
\endfoot
\hline
\endlastfoot

\textbf{Expand access to GenAI for an equitable distribution of benefits} \newline 
Media \cite{Brown_Finney_Korgaonkar_McMillan_Perkins_2023, Pham_Kohli_Llano_Nokuri_Weinstock_2024, Greene-Santos_2024,SimoneNoveck_2025,BlackEnterprise_2025, Abdur-Rahman_2025} \newline 
SLR \cite{Verma2025-wv, Solyst2025-mi, tanksley2025ethics, bosco2025alzheimer, kay2024epistemic} & 
\textit{"As AI usage in education expands, stakeholders should consider how AI tools can become more accessible, particularly in rural and urban communities that have larger populations of low-income students and students of color." \cite{Pham_Kohli_Llano_Nokuri_Weinstock_2024}} \\ \addlinespace

\textbf{Resist GenAI} \newline 
Media \cite{Brown_Finney_Korgaonkar_McMillan_Perkins_2023, Attiah_2025, Greene-Santos_2024, SimoneNoveck_2025, Blay_2025, Berkeley_2025} \newline 
SLR \cite{Xu2025-qt, Basoah2025-yt, dorn2025reinforcing} & 
\textit{"Either way, we need to be ready to resist whatever internet decay Meta has in store for us next." \cite{Attiah_2025}} \\ \addlinespace

\textbf{Regulate GenAI through policy} \newline 
Media \cite{Banks_2025, Brown_Finney_Korgaonkar_McMillan_Perkins_2023,Simpson_2024, Stefflbauer_2024, Pham_Kohli_Llano_Nokuri_Weinstock_2024, Klein_2024, Williams_2024, Greene-Santos_2024,SimoneNoveck_2025, Eledroos_2025, Blay_2025, Berkeley_2025, Marrinan_2025} \newline 
SLR \cite{Elsharif2025-bv, Verma2025-wv,  Webster2025-lw, Xu2025-qt, Nias2025-gf} & 
\textit{"Legal interventions exist and should be used to mandate swift action when a company's technology threatens or unfairly excludes any of us." \cite{Stefflbauer_2024}} \\ \addlinespace

\textbf{Devise actionable strategies to mitigate harm from GenAI} \newline 
Media \cite{Stewart_2024, Brown_Finney_Korgaonkar_McMillan_Perkins_2023, Simpson_2024, Hale_2024, Klein_2024, Williams_2024,Keenan_2023, Greene-Santos_2024, SimoneNoveck_2025} \newline 
SLR \cite{kuck2023generative, ren2024large, maluleke2022studying, cheng2023marked, Wilson2025-oa, Lutz2025-ks, Kim2025-dl, Dinkar2025-hh,  Rastogi2025-ao, Webster2025-lw, Zhou2025-wt,  tanksley2025ethics,  Basoah2025-yt, Proebsting2025-dp, Ocumpaugh2025-ri, Nias2025-gf, Harvey2025-et, bosco2025alzheimer, hassan2025dialectic, mire2025rejected, dorn2025reinforcing, bilalpur2025topic, kay2024epistemic, croskey2025liberatory} & 
\textit{"Warm companies should be thinking now about how to re-skill workers at risk of gen AI displacement. No job is future-proof; the emphasis, for both workers and employers, should be on honing future-proof skills." \cite{Banks_2025}} \\ \addlinespace

\textbf{Reform GenAI through technical intervention} \newline 
Media \cite{Gupta_2024, Brown_Finney_Korgaonkar_McMillan_Perkins_2023, Simpson_2024, Gaskins_2024, Small_2023, Beer_2025, Asare_2023, Pham_Kohli_Llano_Nokuri_Weinstock_2024, Greene-Santos_2024, SimoneNoveck_2025, Fatou_Blackstock_2025, Richards_2025} \newline 
SLR \cite{osti2024collaborating, shankar2024coraal, xu2023harmonyvisage, kuck2023generative, yucer2020exploring, howard2024socialcounterfactuals, zhou2024bias, zhang2023chatgpt, lakkaraju2023can, gupta2024aavenuedetectingllmbiases, lee2024vision, slaughter2023pre, lee2024can, zayed2024fairness, wu2024stereotype, deas2023evaluation, gupta2023bias, nezhad2024fair, groenwold2020investigating, maluleke2022studying, bautista2023health, cheng2023marked, devinney2024we, Verma2025-uy, Junias2025-wl, Narayan2025-bk, Bhardwaj2025-hj, Zhao2025-cw, Elsharif2025-bv, Jung2025-te, Srinivasan2025-dp, Wilson2025-hk, Wilson2025-oa, Sakunkoo2025-rg, Saffari2024-xk, Shukla2025-sm, Malik2025-dp, Yanaka2025-tb, dossou2025towards, Finch2025-rt, Porwal2025-hv, Coggins2025-pt, Kim2025-dl, Wang2025-cn, Verma2025-wv, Khorramrouz2025-qm, Fu2025-js, Kondrup2025-vg, Solyst2025-mi, Swati2025-ai, Shukla2025-bt,  Abbas2025-xi, Rastogi2025-ao, Webster2025-lw, Haider2025-tv, Eschner2025-qy, Lee2025-na, Cantini2025-fp, Zhou2025-wt,  Liang2025-xb, sandoval2025my, Xu2025-qt, tanksley2025ethics, Basoah2025-yt, Proebsting2025-dp, Ocumpaugh2025-ri, Nias2025-gf, Harvey2025-et, Basoah2025-eq, bosco2025alzheimer, morales2025imagebite, dildora2025democratizing, hassan2025dialectic, wan2025white, mire2025rejected, wilson2025bias, kay2024epistemic, croskey2025liberatory} & 
\textit{``To overcome the anti-blackness that is rampant within AI, we must ensure that those programming AI systems are operating from an anti-racist and anti-oppressive lens.'' \cite{Asare_2023}} \\
\end{longtable}
\end{document}